\documentclass[aps,prd,twocolumn,showpacs,superscriptaddress,groupedaddress]{revtex4} 
\usepackage[latin1]{inputenc}
\usepackage{graphicx}
\usepackage{amssymb}
\usepackage{color}
\usepackage{float}
\usepackage{amsmath}
\usepackage{mathrsfs}
\usepackage{mathtools}
\usepackage{amsfonts}
\usepackage{dcolumn}
\usepackage{hyperref}
\usepackage{amsthm}
\usepackage{color}
\usepackage{upgreek}

\def\3nab{\tilde{\nabla}}

\def\be {\begin{equation}}
\def\ee {\end{equation}}
\def\ba {\begin{eqnarray}}
\def\ea {\end{eqnarray}}
\newtheorem{thm}{Theorem}
\newtheorem{prop}{Proposition}

\newcommand{\barray}{\begin{array}}
\newcommand{\earray}{\end{array}}

\hypersetup{
	colorlinks=true,
	citecolor=red,
    linkcolor=blue,
    urlcolor=blue,
	}
\begin{document}

\title{What makes the gravito-thermodynamic arrow of time coincide with that of proper time?}
\author{Samarjit Chakraborty}
\email{samarjitxxx@gmail.com}
\affiliation{Astrophysics Research Centre, Discipline of Mathematics, School of Agriculture and Science, University of KwaZulu-Natal, Private Bag X54001, Durban 4000, South Africa}
\author{Rituparno Goswami}
\email{goswami@ukzn.ac.za}
\affiliation{Astrophysics Research Centre, Discipline of Mathematics, School of Agriculture and Science, University of KwaZulu-Natal, Private Bag X54001, Durban 4000, South Africa}

\begin{abstract}
In the pursuit of a geometrical resolution to this important question for the evolution of spherically symmetric perfect fluids in general relativity, we excavated a novel result: the presence of volume distortions and inhomogeneities are, though necessary, but not sufficient for the total gravitational entropy to monotonically increase with proper time. We need a further refinement of the space of distortions and only those that favour shell crossing caustics, causing local flocculations and structure formations, can ensure the alignment. This result will have far reaching consequences for the study of thermodynamical viability of various gravitational collapse and cosmological solutions of Einstein field equations.

\end{abstract}

\maketitle
In his landmark 1865 paper \cite{Clausius}, Rudolf Clausius formulated the deep results on the first and second laws of thermodynamics: {\it while the energy of the universe is constant, the entropy of the universe tends to a maximum}. These results, when incorporated in the statistical mechanical formalism by Boltzmann, naturally explained why the non-interacting gas molecules evenly spread out inside a container, as that represents the overwhelming majority of all available microstates. This simplified picture, however, breaks down in a cosmological scenario where the cosmological matter field interacts with itself  via the gravitational field. 

The cosmic microwave background indicates that the early universe originated from an exceptionally smooth state. Its near-perfect blackbody spectrum demonstrates an initial state of thermal equilibrium, typically associated with maximum entropy. This does introduce a paradoxical situation: {\it if the second law of thermodynamics holds, how can such a universe evolve further?} The resolution lies in the effect of gravity, that drives structure formation by clumping matter together. This demonstrates, how the standard thermodynamic intuition favouring spatial uniformity, breaks down in the presence of gravitational interaction and therefore highlighting the crucial importance of incorporating the entropy of gravitational field (or the spacetime geometry) in our calculations. 

A parallel challenge emerges in black hole thermodynamics. To resolve the missing entropy problem in this context, Bekenstein proposed that black hole entropy is proportional to its horizon area and formulated the generalised second law of thermodynamics, asserting that the total entropy, i.e., the sum of ordinary matter entropy and horizon entropy must monotonically increase \cite{Bekenstein,Hawking}. Penrose \cite{penrose1,penrose2} addressed this fundamental issue by proposing the Weyl Curvature Hypothesis (WCH), which posits that the free gravitational field carries its own intrinsic entropy, governed by the Weyl curvature tensor. Consequently, gravitational entropy grows alongside cosmic structure formation, establishing a thermodynamic arrow of time.

In general relativity, the independent components of the Riemann curvature tensor $R^{a}{}_{bcd}$ encapsulate the full geometric information of spacetime (and hence the gravitational field). This tensor admits an irreducible decomposition
\begin{equation}
R_{abcd} = G_{abcd} + E_{abcd} + C_{abcd},
\end{equation}
where $G_{abcd}$ and $E_{abcd}$ are algebraically determined by the matter energy-momentum tensor $T_{ab}$ via the Einstein field equations. Consequently, these two trace terms vanish in vacuum. The trace-free Weyl tensor $C_{abcd}$, however, remains non-zero, representing the free gravitational field and purely geometric degrees of freedom. This makes the Weyl tensor the natural geometrical structure for defining gravitational entropy.

In the pursuit of defining a gravito-thermodynamic arrow of time, proposals by Bonnor and others \cite{Bonnor1,Bonnor2,Bonnor3,arrow1,arrow2,arrow3} relied primarily on pure curvature invariants, but these heuristic measures lack a direct connection to fundamental thermodynamics. In this letter, we argue that defining a physical gravito-thermodynamic arrow of time, based on the gravitational entropy as defined by Clifton, Ellis and Tavakol \cite{CET} would give a more transparent picture, from the point of view of a relativistic thermodynamic framework. Given that matter and gravitational entropy are additive, we propose tracking the local evolution of the total point-wise entropy density `$s$' within a finite spatial volume `$V$', rather than relying solely on geometric scalar ratios. This leads to our central inquiry:\\

\textbf{Question:} \textit{What makes the gravito-thermodynamic arrow of time coincide with that of proper time?}\\

To address this question, for a spherically symmetric configuration, we adopt the semi-tetrad $1+1+2$ covariant decomposition in Locally Rotationally Symmetric class II (LRS-II) spacetimes \cite{H8,H9,H10,H11,H12,H13,H14}, which encompasses all spherically symmetric geometry. LRS-II spacetimes are irrotational and possess a continuous, non-trivial spatial isotropy group at every point, defining a unique, covariantly preferred spatial direction $e^a$. Selecting a unit timelike vector $u^a$ (preferably along the matter worldlines, in the presence of matter) and using the preferred orthogonal spacelike unit vector $e^a$, the metric decomposes as \cite{R18}
\begin{equation}\label{decomp}
g_{ab} = -u_au_b +h_{ab}=-u_a u_b + e_a e_b + N_{ab},
\end{equation}
where $h_{ab} \equiv g_{ab} + u_a u_b$ projects onto the 3-space orthogonal to $u^a$, and $N_{ab} = h_{ab} - e_a e_b$ projects onto the 2-spherical shells. Kinematical quantities associated with the timelike congruence $u^a$ are  the expansion scalar ($\Theta\equiv \nabla_au^a$), the 3-acceleration vector ($\dot{u}^a\equiv u^b \nabla_b u^a$) and the shear 3-tensor $\sigma_{ab}$, which is the projected symmetric trace free part of $\nabla_au_b$. The Weyl curvature tensor decomposes into the electric part ($E_{ab} \equiv C_{acbd}u^c u^d$), describing tidal forces and inhomogeneities, while the magnetic part vanishes identically in spherical symmetry. Similarly, the matter energy-momentum tensor $T_{ab}$ decomposes as
\begin{equation}\label{EMT}
T_{ab} = \mu_m u_a u_b + p_m h_{ab} + 2u_{(a}q_{b(m))} + \pi_{ab(m)},
\end{equation}
where $\mu_m$, $p_m$, $q_{a(m)}$, and $\pi_{ab(m)}$ denote energy density, isotropic pressure, heat flux, and anisotropic stress, respectively. Projecting these 3-tensors and 3-vectors along $e^a$ reduces the spatial degrees of freedom to a minimal set of covariant scalars: spatial expansion of $e^a$ ($\phi\equiv h^a_ch^c_b\nabla_ae^b$),  acceleration scalar ($\mathcal{A} \equiv \dot{u}^a e_a$), shear scalar ($\Sigma \equiv \sigma_{ab}e^a e^b$), electric Weyl scalar ($\mathcal{E} \equiv E_{ab}e^a e^b$), heat flux scalar ($Q_m \equiv q_{a(m)} e^a$), and anisotropic stress scalar ($\Pi_m \equiv \pi_{ab(m)}e^a e^b$). Consequently, LRS-II spacetimes are completely characterized by the set of scalars
\begin{equation}\label{set1}
\mathcal{D} \equiv \left\{\Theta, \mathcal{A}, \Sigma, \phi, \mathcal{E}, \mu_m, p_m, \Pi_m, Q_m \right\},
\end{equation}
and their evolution along $u^a$ (denoted by a dot, $\dot{} = u^a \nabla_a$) and spatial propagation along $e^a$ (denoted by a hat, $\hat{} = e^a \nabla_a$) fully specifying the Ricci and Bianchi identities.

To calculate the relativistic entropy evolution of the matter, we follow the seminal work by Ehlers \cite{Ehlers}. The conservation equations for the matter field is given via the divergence free property of the Energy Momentum tensor ($\nabla_b T^{ab}=0$), which when contracted with $u^a$ gives
 \begin{equation}\label{contra}
\dot{\mu}_m+(\mu_m + p_m)\Theta +\pi_{ab(m)}\sigma^{ab}+ \nabla_a q_{(m)}^a +q_{a(m)} \dot{u}^{a}=0.
\end{equation}
Now for continuous bulk matter, we have an uniquely determined local 4-velocity $u^a$ (which is called ``{\it local Barycentric velocity}"\cite{Meixner}) such that, the main part of the energy density ``$\mu_{m}$" is the positive rest mass proper density $\rho$, that remains conserved ($\nabla_a (\rho u^a)=0$), {\it i.e.} $\dot{\rho}+\Theta \rho =0$. Now defining a ``specific volume" $V\equiv 1/\rho$ (the volume that contains unit mass) we get $\rho \dot{V}=\Theta$.
Introducing the specific internal energy $U_m \equiv \mu_m / \rho$, the rate of change of the energy density becomes
\begin{equation}\label{dotmu}
\dot{\mu}_m=\dot{U}_m \rho-\mu_m \Theta.
\end{equation}
Substituting the definition of specific volume and \eqref{dotmu} into Eq. \eqref{contra} yields the local energy balance equation
\begin{equation}
\rho\left[\dfrac{dU_m+p_m dV}{d\tau}\right]=-\left(\pi_{ab(m)} \sigma^{ab}+ \nabla_a q_{(m)}^a +q_{a(m)} \dot{u}^a\right),
\end{equation}
where $\tau$ denotes the proper time along these Barycentric world-lines. Also along these worldlines, the local thermodynamic state obeys the Gibbs relation $T ds_m = dU_m + p_m dV$ (where $T$ is the absolute temperature of the matter field), and therefore the time rate of change of matter entropy $s_m$ along the flow lines satisfies the following
\begin{equation}
T\dfrac{ds_m}{d\tau}\equiv T\dot{s}_m=-V\left(\pi_{ab(m)} \sigma^{ab}+ \nabla_a q_{(m)}^a +q_{a(m)} \dot{u}^a\right).
\end{equation}

From the above equation, it is clear that perfect fluids are especially interesting, because in relativistic thermodynamics, the matter entropy $s_{m}$ remains constant along the flow, i.e., $\dot{s}_m=0$ \cite{Roy, Ehlers} and hence the evolution is {\it isentropic}. Therefore, the only source of entropy production comes from the gravitational entropy ($\dot{s}_{grav}$), thereby providing a gravito-thermodynamic arrow of time. The condition $\dot{s}_{grav}>0$ ensures that  this arrow coincides with that of proper time. For the rest of our analysis, to ensure better transparency and physical viability, we will only consider a perfect fluid form of matter that obeys physically realistic energy conditions.

In order to obtain $s_{\text{grav}}$, we need to first have a properly thermodynamically motivated gravitational entropy that has all the desirable properties. It was shown that the Bel-Robinson tensor $T_{abcd}=\frac14\left(C_{eabf}C^{e~~~f}_{~cd}+C^*_{eabf}C^{*e~~~f}_{~cd}\right)$ is a perfect candidate and a unique Maxwellian tensor, that acts as the `super energy momentum' tensor for free gravitational fields \cite{E7}. With a caveat, of course, that this tensor has dimensions of $L^{-4}$, rather than the required $L^{-2}$ . This dimensionality issue was addressed by Bonilla and Senovilla \cite{sen1}, who showed that for this symmetric and tracefree four-index tensor, one can define a symmetric 2-index square root, which is  unique In Petrov type D and N spacetimes. Subsequently, making use of this property and imposing conservation in vacuum, Clifton, Ellis and Tavakol proposed a mechanism to obtain the entropy of free gravitational fields \cite{CET}, by defining this 2-index square root ($\mathscr{T}_{ab}$) as the energy momentum tensor for the free gravity.  This mechanism successfully reproduced the Bekenstein-Hawking  entropy of a static black hole and has been thoroughly investigated in cosmology \cite{CET}. Moreover, using this, one can reconstruct the black hole entropy from a stellar collapse in both static and radiative spacetimes \cite{Aqua,Shamima}. Most importantly, it connects the Weyl curvature hypothesis to the relativistic thermodynamics framework.  

To perform a thermodynamical analysis of free gravity, exactly analogous to matter fields will cause a problem, as we cannot define rest mass density for a free gravitational field. However, this problem can be circumvented if {\it we a-priory choose the  timelike unit vector $u^a$, which should be the matter Barycentric vector in presence of matter, but for vacuum spacetimes it is arbitrary}. Given the expansion scalar as $\Theta=\nabla_a u^a$, we then can define a ``specific volume" $V$ by the relation
\begin{equation}
\Theta=\dfrac{u^a \nabla_a V}{V}\,.
\end{equation}
This will then naturally enable us to define the specific internal energy of the free gravitational field as $U_{grav}\equiv \mu_{grav}V$.
Once $u^a$ is defined, we can decompose the effective energy momentum tensor of the free gravitational field  $\mathscr{T}_{ab}$ exactly analogously
\begin{equation}
\mathscr{T}_{ab}=\mu_{grav}u_a u_b +p_{grav}h_{ab}+2 u_{(a}q_{b(grav))}+\pi_{ab(grav)}.
\end{equation}
The free gravity energy momentum tensor is {\it not divergence free} in presence of a matter field, and hence we define a  "{\it gravitation energy current}" $J^a\equiv -\nabla_b \mathscr{T}^{ab}$ and contracting this with $u^a$ we obtain
\begin{align}\label{Gibbs1}
\frac{{dU}_{grav} + p_{grav}{dV}}{d\tau}= & V \Big[ J^a u_a -\nabla_a q_{(grav)}^a - \dot{u}^{a} q_{a(grav)}  \nonumber\\
& -\sigma_{ab}\pi^{ab}_{(grav)} \Big].
\end{align}
Assuming that Gibbs relation holds for free gravity, we can write the above equation as
\begin{align}
T_{grav} \dfrac{ds_{grav}}{d\tau}= & V\Big[J^a u_a -\nabla_a q_{(grav)}^a - \dot{u}^{a} q_{a(grav)} \nonumber\\
& -\sigma_{ab}\pi^{ab}_{(grav)}\Big].
\end{align}
In a general Petrov type D spacetime, the energy momentum tensor for the free gravitational field is given by \cite{CET}
\begin{equation}
\mathscr{T}_{ab}=\sqrt{\dfrac{2W}{3}}\Big\{2(u_{a}u_{b}-e_{a}e_{b})+N_{ab}\Big\},
\end{equation}
where, for spherical symmetry we have $W=3{\mathcal{E}^2}/8$. Now, using the spacelike preferred direction $e^a$, we can identify the following analogous scalars  
$ \mu_{grav}=\mathcal{E},\,\,p_{grav}=0,\,\,\Pi_{grav}=-\mathcal{E},\,\,Q_{grav}=0$.
Therefore, the entropy evolution of the free gravitational field becomes
\begin{equation}\label{ther}
T_{grav}\dot{s}_{grav}=-V\left[\frac{3}{2}\Sigma\Pi_{grav}- J^au_a\right]
\end{equation}
From, the singly contracted Bianchi Identities, we have  \cite{GOSNewFron}
\begin{equation}
- J^au_a=u_a \nabla_b \mathscr{T}^{ab}=-\dot{\mathcal{E}}+\frac{3}{2}\mathcal{E}\left(\Sigma-\frac{2\Theta}{3}\right).
\end{equation}
The above equation can be simplified through the following scalar equations derived from singly and doubly contracted Bianchi Identities  \cite{GOSNewFron}:
\begin{align}\label{evol1}
\dot{\mathcal{E}}-\frac{\dot{\mu}_m}{3}=&\left(\Sigma-\dfrac{2\Theta}{3}\right)\left(\dfrac{3}{2}\mathcal{E}+\frac{\mu_m+p_m}{2}\right),\\
\dot{\mu}_m=&-\Theta(\mu_m+p_m)\,,\label{evol2}
\end{align} 
to get 
\begin{equation}\label{weyl}
u_a \nabla_b \mathscr{T}^{ab}=\dfrac{\Sigma}{2}(\mu_m+p_m).
\end{equation}
Consequently, in presence of a perfect fluid, the thermodynamic identity for the free gravitational field becomes
\begin{equation}\label{sdot}
T_{grav}\dot{s}_{grav}=-\frac{3V\Sigma}{2}\left[-\mathcal{E}+\dfrac{(\mu_m+p_m)}{3}\right]
\end{equation}
Note that the above expression is the 1+1+2 version of the obtained expression (Eq. 24) in \cite{CET}.  This expression highlights the importance and the role of spacetime volume deformation (in terms of shear) for the time evolution of gravitational entropy and brings us to the following important result:
\begin{prop}
Spherically symmetric shear free perfect fluid spacetimes are isentropic with respect to the total entropy, where both the matter entropy $s_m$ and the gravitational entropy $ {s}_{grav}$ are conserved separately.
\end{prop}
This also corroborates the findings of Herrera \cite{Herrera}, that shear-free perfect fluid spacetimes are not thermodynamically viable, as the total entropy doesn't increase with time.

Although we don't need an exact expression of the gravitational temperature $T_{grav}$, for the sake of completeness we mention the temperature function that is inspired from the black hole thermodynamics and quantum field theory in curved spacetime \cite{CET,Shamima}. Recently, it was studied geometrically in \cite{affine} and found to be the non-affinity factor of the local radial Newman-Penrose null geodesics in the LRS-II class spacetimes. In terms of the kinematical scalars,  this gravitational temperature function is given by the following expression
\begin{equation}
T_{grav}=\frac{1}{2\pi}\vert\mathcal{A}+\frac{\Theta}{3}+\Sigma\vert.
\end{equation}

To transparently obtain the sign of $\dot{s}_{grav}$ in Eq. (\ref{sdot}), we use the expression of Misner-Sharp mass $\mathcal{M}$, that gives the total gravitational mass enclosed within a spherical $2$-shell. In terms of the covariant scalars this is expressed as \cite{GOSNewFron}
\begin{equation}\label{Mass}
\mathcal{M} =\frac{1}{2\sqrt{K}}\left[1-\frac{\nabla_{a}K \nabla^{a}K}{4K^3}\right] =\frac{1}{2K^{\frac{3}{2}}}\left(\frac{1}{3}\mu_{m} - \mathcal{E}\right).
\end{equation}
where $K$ is the Gaussian curvature of the spherical 2-shells (inverse square of the area radius $\mathcal{R}$), and is given as 
\begin{equation}
K\equiv\frac{1}{\mathcal{R}^2}=\frac{1}{3}\mu_{m} -\mathcal{E}+\frac{1}{4}\left[\phi^2-\left(\Sigma-\frac{2}{3}\Theta\right)^2\right].
\end{equation}
We can also obtain the directional derivatives of $K$ along $u^a$ and $e^a$ as
\begin{equation}
\label{Kdothat}
\dot{K} =\left(\Sigma - \frac{2}{3}\Theta \right)K\;\;;\;\; \hat{K} =  -\phi K.
\end{equation}
Therefore, for a properly oriented gravito-thermodynamic arrow of time, the following condition needs to be satisfied
\begin{equation}\label{arrowpf}
T_{grav}\dot{s}_{grav}=-\dfrac{V\Sigma}{2} \left[\frac{6\mathcal{M}}{\mathcal{R}^3}+p_m \right]>0.
\end{equation}

The first term in the square bracket above is strictly non-negative and the term $\mathcal{M}/\mathcal{R}^3$ denotes some measure of average gravitational energy density within a given shell. Consequently, if the matter field obeys all the reasonable energy conditions, it can be safely assumed that the matter pressure will be in the range $\vert p_m\vert<\mathcal{M}/\mathcal{R}^3$, thus making the square parenthesis term non-negative. This then leads us to the following theorem:
\begin{thm}
For the evolution of spherically symmetric perfect fluids obeying all reasonable energy conditions, the  gravito-thermodynamic arrow of time is aligned with that of the proper time along a worldline iff the covariant shear scalar $\Sigma$ stays strictly negative on the worldline.
\end{thm}

It is quite intriguing to see that not only the {\it presence} of volume distortion, but a special subset of the space of all possible volume distortions that makes the shear scalar strictly negative, is the reason for the alignment of the arrows of time. Here comes the next important question:\\

\textbf{Question:} \textit{How do we physically interpret the negative shear scalar?}\\

A priory, it is important to note that the {\it sign of shear scalar doesn't interfere with the gravitational focussing}, as the Raychaudhuri equation depends on $\Sigma^2$ and the same goes with the Friedmann constraint in the cosmological setting.  A well known spacetime which is vacuum, spatially homogeneous and where the natural worldlines have negative shear scalar throughout, is the interior of a Schwarzschild black hole, where the metric is given as \cite{LargeScale}, 
\begin{equation}
ds^2=-\frac{dt^2}{\left(\dfrac{2m}{t}-1\right)}+\left(\dfrac{2m}{t}-1\right)dr^2+t^2d\Omega^2\;.
\end{equation}
It can be easily checked that for the congruence $u^a=\sqrt{\dfrac{2m}{t}-1}\delta^a_t$, the shear scalar is
\begin{equation}
\Sigma=-\frac23\frac{3m-t}{t\sqrt{t(2m-t)}}
\end{equation}
and remains strictly negative in the allowed interval $t\in(0,2m)$. Thus, for a test shell falling in the black hole, the entropy increases in the interior region till the shell hits the singularity and in the exterior region this is manifested by increase in the horizon area $\delta A_H>0$, as predicted by the black hole thermodynamics and thus {\it pointing towards the classical stability of black holes}. 

A more illuminating example, that clearly differentiates the effects of positive and negative shear scalar, would be Lemaitre Tolman Bondi (LTB) inhomogeneous dust spacetimes. For transparency, let us consider the gravitational collapse of a marginally bound dust like matter \cite{Ritu}, where the metric is given as
\begin{equation}
ds^2=-dt^2+(\mathcal{R}_{,r})^2dr^2+\mathcal{R}^2d\Omega^2\,,
\end{equation}
with the area radius $\mathcal{R}(t,r)=ra(t,r)$. The inhomogeneous scale factor $a(t,r)\equiv(1-\sqrt{m(r)}\,t)^{(2/3)}$ is scaled in a way such that at the initial epoch ($t=0$) $a(0)=1$ and at the singularity of a given shell $a=0$. The function $m(r)$ can be described in terms of the initial density profile as $m(r)=r^{-3}\int_0^r \mu_0(r)r^2dr$. The shear scalar for this spacetime is given by
\begin{equation}\label{nsigma}
\Sigma=\frac{2}{3}\left(\dot{\ln[\mathcal{R}_{,r}]} - \dot{\ln[\mathcal{R}]}\right)\,.
\end{equation}
A straightforward calculation reveals that constraining the shear scalar to be negative, would necessarily imply $ m(r)_{,r}>0$ ({\it i.e.} the function increases as we move away from the centre). Interestingly, this is the necessary and sufficient condition for the presence of {\it Shell crossing singularities} in this spacetime, as the outer shells reach the zero-area radius singularity faster than the inner shells (which could have been eliminated by a positive shear scalar). Shell crossing  singularities occur when nearby shells of matter intersect, creating momentary density divergences where some of the curvature scalars could blow up \cite{psjbook1,psjbook2}. This may happen at any stage of the collapse and the situation is very similar to the formation of {\it caustics} in Newtonian fluid dynamics where two nearby fluid layers cross each other and thereby create shock waves travelling through the medium. In the context of general relativity, these shell crossing singularities are known to be weak in nature which are resolved through a suitable extension of the spacetime. The presence of these singularities strongly indicates local structure formation where the usual dustlike approximation breaks down. A similar effect can be seen, if we consider expanding LTB models as well. 

The above phenomenon is by no means an artefact of dustlike matter only. The same effect continues for other spherically symmetric inhomogeneous and shearing perfect fluid spacetimes which are spatially inhomogeneous with $\phi>0$. From the propagation equation  \cite{GOSNewFron,supp} 
\begin{equation}
\left(\hat{\Sigma}-\frac{2\hat{\Theta}}{3}\right)=-\frac{3}{2}\phi \Sigma\,,
\end{equation}
it is clear that a negative shear scalar will force the normalised rate of change of Gaussian curvature of a given shell ($\dot{\ln[K]}$) to increase as we go along the preferred spatial direction. For expanding scenario ($\dot{K}<0$), this will imply that the inner shells will cross the outer ones at some point triggering the shell crossing singularities. A similar phenomenon would happen for collapsing scenario  ($\dot{K}>0$) as well where the outer shells will cross the inner ones. Hence in all these  spacetimes a negative shear creates structures with increasing gravitational entropy.

To reaffirm, we know that there is an intimate relationship between the volume distortion (represented by the shear of the worldlines) and inhomogeneities (caused by the tidal effects due to the electric Weyl) \cite{PSJRMND, GOSNewFron}. Both shear and electric part of the Weyl sources each other in their respective evolution equations. One would then expect that presence of these effects will suffice for the increase in the total entropy for an evolving perfect fluid along a given wordline, as both distortion and inhomogeneities are measures of disorderliness. However the novel result that emerged for our analysis tells us more: {\it only the presence of volume distortion and tidal effects are not enough}. We need a further refinement in the set of all possible distortions, and choose only those that necessarily favours shell crossing caustics (and thus favouring local focculation and structure formation), to make the total entropy and disorderliness of the evolving fluid increase along the worldline and thus aligning the arrows of the gravito-thermodynamic time with that of the proper time. The next obvious question would be to check the effects of non-sphericity (specifically rotation) within Petrov Type D spacetimes. We keep this for future investigations. 

 \section{Acknowledgements}

SC thanks University of KwaZulu-Natal (UKZN) for Postdoctoral fellowship. RG thanks UKZN and the National Research Foundation (NRF), South Africa for support.

\end{document}